\documentclass[12pt,draftcls,onecolumn,letterpaper]{IEEEtran}
\usepackage{amsmath,amssymb,eucal,graphicx}
\usepackage{epsfig}
\usepackage{exscale}
\usepackage{color}
\usepackage{psfrag}

\def\LB{\left(}         
\def\RB{\right)}        

\def\tfb{T_{\rm fb}}
\def\nt{N_t}

\newfont{\bbb}{msbm10 scaled 500}

\newfont{\bb}{msbm10 scaled 1100}
\newcommand{\CC}{\mbox{\bb C}}

\newcommand{\EE}{\mbox{\bb E}}

\newcommand{\hv}{{\bf h}}

\newcommand{\sv}{{\bf s}}

\newcommand{\vv}{{\bf v}}
\newcommand{\xv}{{\bf x}}

\newcommand{\zv}{{\bf z}}

\newcommand{\Hm}{{\bf H}}

\newcommand{\Cc}{{\cal C}}

\newcommand{\Kc}{{\cal K}}
\newcommand{\Lc}{{\cal L}}

\newcommand{\Nc}{{\cal N}}

\newcommand{\Sc}{{\cal S}}

\newcommand{\eqdef}{\stackrel{\Delta}{=}}

\DeclareFontFamily{U}{cmfi}{}
\DeclareFontShape{U}{cmfi}{m}{n}{ <-> cmfi10 }{}
\DeclareSymbolFont{CMFI}{U}{cmfi}{m}{n}

\newcommand{\herm}{{\sf H}}

\begin{document}

\vspace{-3cm}
\title{Training and Feedback Optimization for Multiuser MIMO Downlink}

\author{Mari Kobayashi$^1$, \qquad
%
Nihar Jindal$^2$,  \qquad
Giuseppe Caire$^3$
}
\maketitle
\noindent
{\small $^1$ SUPELEC, Gif-sur-Yvette, 91192, France \\
$^2$ University of Minnesota,  Minneapolis MN, 55455 USA \\
$^3$ University of Southern California, Los Angeles CA, 90089 USA}

\begin{abstract}

We consider a MIMO fading broadcast channel where
the fading channel coefficients are constant over time-frequency blocks that
span a coherent time $\times$ a coherence bandwidth. In closed-loop systems, channel state information at transmitter (CSIT) is acquired by the downlink training sent by the base station and an explicit feedback from each user terminal. In open-loop systems,
CSIT is obtained by exploiting uplink training and channel reciprocity. We use a tight closed-form lower bound on the ergodic achievable rate in the presence of CSIT
errors in order to optimize the overall system throughput, by taking explicitly into account the
overhead due to channel estimation and channel state feedback.
Based on three time-frequency block models inspired by actual systems, we provide some useful guidelines for the overall system optimization. In particular, digital (quantized) feedback is found to offer a substantial advantage over analog (unquantized) feedback.
\end{abstract}

{\bf Keywords:} MIMO broadcast channel, Multiuser MIMO Downlink, Channel State Information Feedback, Channel Estimation.


\section{Introduction}\label{sect:introduction}

The downlink of a wireless system with one Base Station (BS) with $\nt$ antennas and $K$ User Terminals (UTs)
with a single antenna each is modeled by a MIMO Gaussian broadcast channel \cite{caire2003atm}, defined by
\begin{equation} \label{model}
y_k[i] = \hv_k^\herm \xv[i] + z_k[i], \;\; k = 1,\ldots,K
\end{equation}
for $i = 1,\ldots, T$,  where $y_k[i]$ is the channel output
at UT $k$, $z_k[i] \sim \Cc\Nc(0,N_0)$ is the corresponding
Additive White Gaussian Noise (AWGN) process, $\hv_k \in \CC^{\nt}$ is the vector of channel coefficients
from the BS antenna array to the $k$-th UT antenna and $\xv[i]$ is the
vector of channel input symbols transmitted by the BS, subject to the average power constraint
$\EE[|\xv[i]|^2] \leq P$ (enforced for each channel use $i$). We denote the downlink signal-to-noise ratio (SNR) at each UT by $\rho \triangleq \frac{P}{N_0}$.
We assume a block fading model where the channel vectors $\{\hv_k\}$ remain constant over a coherence block of $T$ channel uses.
The block length $T$ is related to two physical channel parameters, the coherence time $T_c$ and the coherence bandwidth $W_c$
by $T = W_c T_c$. For example, taking as typical values $W_c = 500$ kHz and $T_c = 2.5$ ms (from \cite{tse2005fwc}),
we obtain $T = 1250$ channel uses.

Albeit suboptimal, zero-forcing (ZF) beamforming with $K=\nt$ users
captures the fundamental trend in terms of degrees of freedom (or ``multiplexing gain'')
\cite{tse2005fwc}.
Therefore, we focus on this case for its analytical tractability.
In order to perform ZF beamforming (or any other multiuser MIMO precoding), the BS must have an accurate
estimate of the downlink channel. Such information, referred to as the {\em Channel State Information at the Transmitter} (CSIT)
is acquired by using downlink training and
channel state feedback. On the one hand, in TDD systems with self-calibrating devices,
owing to the fact that uplink and downlink take place in the same channel coherence bandwidth,  CSIT can be acquired directly from the
uplink pilot symbols. On the other hand, the uplink-downlink channel reciprocity does not hold in Frequency-Division Duplexing (FDD) systems where uplink and downlink take places in different widely separated frequency bands. This is also the case in Time-Division Duplexing (TDD) systems where uplink and downlink may time-share the same band but the non-linear devices are not self-calibrated and therefore 
induce non-reciprocal effects. In the latter case, an explicit CSIT feedback must be used.
In any case, the rates achievable with ZF beamforming depend critically on the quality
of the CSIT, however, high quality CSIT can be achieved by dedicating a significant amount of time resource 
to downlink training and (for FDD) to channel state feedback. It follows that there is a non-trivial
tradeoff between the benefits of improving the CSIT and the overhead in channel estimation and feedback.

\begin{figure}[n]
    \centering
\includegraphics[width=0.6\columnwidth]{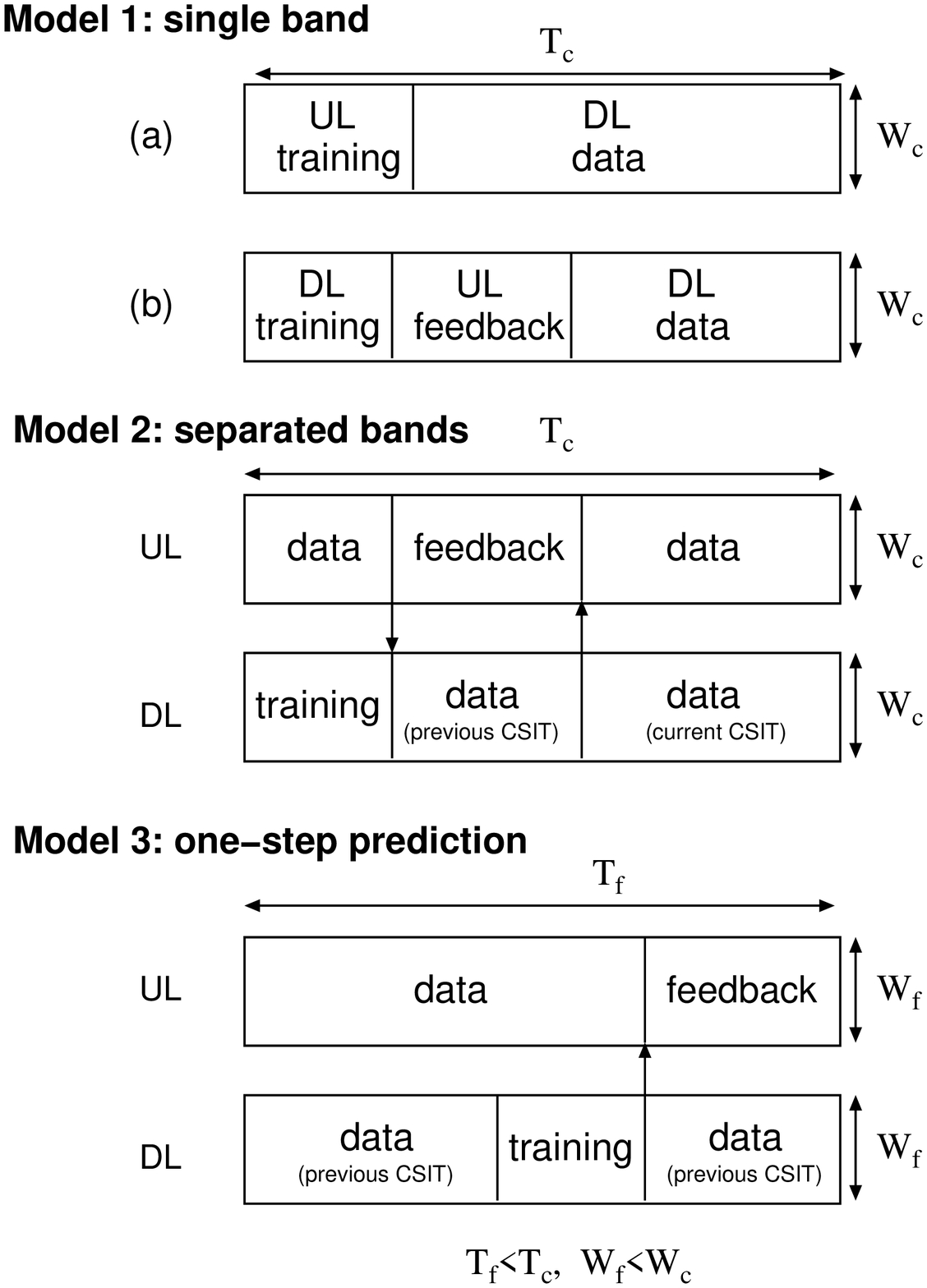}\\
    \vspace{-1em}
    \caption{Different time-frequency block models.}
    \label{fig:Model}
\end{figure}

In this work, we determine the optimum fraction of resources that should
be dedicated to training/feedback in several cases of interest. In particular, we consider three time-frequency block models depicted in Fig. \ref{fig:Model}. These models can be viewed as an idealization of the actual systems such as LTE \cite{sesia2009lte}
and aim at capturing the essential features.
In Section \ref{sect:Optimization}, we consider the optimization of the net spectral efficiency based on model 1 where both training and feedback consume ``downlink'' channel uses.
This analysis applies naturally to TDD with or without reciprocity and FDD where downlink training and (uplink) feedback are performed in the same fading coherence block, via some hand-shaking protocol.
In Section \ref{sec:separate-bands}, we consider a different viewpoint based on models 2 and 3, in which the CSIT feedback consumes ``uplink'' channel uses. These models are more relevant to
FDD systems. The question that we address is ``how much uplink
resource should one pay in order to achieve a certain downlink spectral efficiency?''.
By solving the corresponding optimization problem, we characterize the uplink/downlink spectral efficiency region.
At which point of this tradeoff region the system should operate is a function of the specific system requirements such as uplink/downlink traffic demands.
For a fixed demand, the optimal operation point can be found adaptively.
Further, we study the effect of temporally correlated fading channels and feedback delay where CSIT is obtained through a one-step prediction model (model 3). This corresponds to the case when the downlink block bandwidth $W_f$ and the block length $T_f$ are significantly shorter than the coherence bandwidth $W_c$ and the coherence interval $T_c$, respectively.
Finally, Section \ref{sec:many-users} presents some
considerations for the case of $K > \nt$ users with some downlink scheduling and user selection \cite{YooGoldsmith06}.
This case is very relevant in practice, but its analysis has escaped so far a full closed-form characterization.
Therefore, we provide results by combining Monte Carlo simulation and closed-form analysis.

The optimization of training has been studied
in the context of point-to-point MIMO channels in the literature, e.g.,
\cite{hassibi-hochwald-03it, dana-sharif-hassibi06, Love_SP05,
santipach-honig-isit06, santipach-honig-wcnc07}. In
\cite{hassibi-hochwald-03it}, the point-to-point MIMO communication is
considered and only downlink training is addressed for the case of no CSIT and imperfect Channel State Information at the Receiver (CSIR).
On the other hand, in \cite{Love_SP05}, perfect CSIR is
assumed and the resources to be used for channel feedback are
investigated. In \cite{santipach-honig-isit06,santipach-honig-wcnc07} the model of \cite{hassibi-hochwald-03it} is
extended to also incorporate quantized channel feedback and
transmitter beamforming.  Although the setup is quite similar to ours, the emphasis of \cite{santipach-honig-isit06,santipach-honig-wcnc07}
on the asymptotic regime, where the number of
antennas and $T$ are simultaneously taken to infinity, leads to
rather different conclusions as compared to the present work. In
\cite{dana-sharif-hassibi06},  a MIMO broadcast with downlink training
and perfect channel feedback (i.e., the BS is also able to view the
received training symbols) is considered.  It is shown that the sum
rate achievable with a dirty paper coding-based strategy has a very similar form
to the achievable rate expressions in \cite{hassibi-hochwald-03it}, and thus many of the conclusions from
\cite{hassibi-hochwald-03it} directly carry over. On the other hand,
we consider the more practical case where there is imperfect feedback from each UT to the BS and also study achievable
rates with ZF beamforming, which has lower complexity than dirty-paper coding.
The present work is an extension of \cite{kobayashi2008much,kobayashi2009itw}, where the
same optimization was investigated assuming that both downlink training and uplink feedback are performed within the same block (model 1). In this paper, we provide more complete guidelines on the overall system optimization for the various scenarios of interest.

\section{Channel State Estimation and Feedback}\label{sect:TrainingFB}
When the multiuser MIMO downlink is operated in a closed-loop mode, the CSIT is obtained through the following phases: \\
1) Common downlink training: $T_{\rm tr}$ shared pilot symbols (i.e.,
$\frac{T_{\rm tr}}{\nt}$ pilots per BS antenna) are transmitted on each channel coherence block
to allow all UTs to estimate their downlink channel vectors $\{\hv_k\}$ based on the observation
\begin{equation} \label{training-phase-1-rx}
\sv_k = \sqrt{\frac{T_{\rm tr}P}{\nt}}\ \hv_k + \zv_k.
\end{equation}
Using linear MMSE estimation, the per-coefficient estimation error variance is given by
\begin{eqnarray} \label{csit-error-var}
\frac{1}{1 + \left(\frac{T_{\rm tr}}{\nt}\right) \rho}
\end{eqnarray}
2) Channel feedback: Each UT feeds back its channel estimation
immediately after the training phase.  We focus on the scenario where the feedback channel is modeled
as an AWGN channel with the SNR $\rho$, identical to the nominal
downlink SNR.  Because UT's are assumed to access the feedback
channel orthogonally, a total of $\tfb$ channel symbols translates into
$\frac{\tfb}{\nt}$ feedback channel uses per UT.  Different feedback strategies are described in Section \ref{sect:Optimization}.

The BS obtains the channel state matrix $\widehat{\Hm} = [\widehat{\hv}_1,\ldots, \widehat{\hv}_{\nt}]$
based on the training/feedback information. Errors in the CSIT available to the BS stems from two sources: the channel estimation error during the common training phase, and the distortion incurred during the feedback phase. Then, the BS computes the ZF beamforming vector $\widehat{\vv}_k$ to be a unit-norm vector orthogonal to the subspace
$\Sc_k= {\rm span}\{\widehat{\hv}_j^\herm : j\neq k\}$ for all $k$.
In this case,  the ergodic rate achievable by UT $k$ with equal-power allocation across UT's and
Gaussian random coding is given by:
\begin{eqnarray} \label{eq-rate}
R_k = \mathbb{E} \left[ \log \left(1 + \frac{ |\hv_k^\herm \widehat{\vv}_k|^2 \frac{\rho}{\nt} }
{1 + \frac{\rho}{\nt} \sum_{j \ne k} |\hv_k^\herm \widehat{\vv}_j|^2 } \right) \right],
\end{eqnarray}
assuming each UT is aware of its received signal-to-interference plus noise ratio (SINR).\footnote{Such knowledge can be
acquired through an additional dedicated training round as discussed in \cite{Submitted}.
This training round does not significantly affect the present work, and thus
is ignored for the sake of simplicity.}
The residual interference due to non-zero  ``leakage''  coefficients $\{|\hv_k^\herm \hat{\vv}_j|\}$
decreases the achievable rate. In \cite{Submitted}, it is shown that the rate in (\ref{eq-rate}) is tightly lower-bounded by
\begin{eqnarray} \label{eq-rate2}
R_k \geq R^{\rm ZF}_k - \overline{\Delta R}_k
\end{eqnarray}
where $R^{\rm ZF}_k$ is the rate achievable with perfect CSIT and $\overline{\Delta R}_k$ denotes the {\em rate gap},
given in closed form by
\begin{eqnarray}
\overline{\Delta R}_k \triangleq \log \left(1 + \frac{\rho}{\nt} \sum_{j \ne k}
\mathbb{E} \left[  |\hv_k^\herm \hat{\vv}_j|^2 \right] \right).
\end{eqnarray}
Assuming that the channel statistics are symmetric over users and space,
$R_k$, $R^{\rm ZF}_k$ and $\overline{\Delta R}_k$ do not depend on $k$, therefore the subscript $k$ will be omitted in the following.
The rate gap depends on $T_{\rm tr}$, $\tfb$ and the training/feedback strategy and will be generally
denoted by the function $\overline{\Delta R}(T_{\rm tr}, \tfb)$.  Explicit expressions are found in \cite{Submitted} for the cases addressed in this paper.

\section{Joint Optimization of Training and Feedback} \label{sect:Optimization}

In this section, we focus on model 1 of Fig. \ref{fig:Model} where training and CSIT feedback consume downlink channel uses. Model 1 (a) refers to the TDD system exploiting the channel reciprocity, while model 1 (b) refers to either the TDD without reciprocity or the FDD system in which the downlink training and the feedback are performed in the same fading coherence block.
In both cases, the maximization of the \textit{net} downlink spectral efficiency is formulated as
\begin{equation} \label{eq-mainopt}
\max_{T_{\rm tr}, \tfb: T_{\rm tr} + \tfb \leq T}     \left(1 - \frac{T_{\rm tr} + \tfb}{T}\right) \left( R^{\rm ZF} - \overline{\Delta R}(T_{\rm tr}, \tfb) \right).
\end{equation}
It is convenient to consider the maximization in two steps, by writing:
\begin{equation} \label{eq-mainopt2step}
\max_{T_t \leq T} ~ \max_{T_{\rm tr} + \tfb = T_t}   \left(1-\frac{T_{\rm tr} + \tfb}{T}\right) \left( R^{\rm ZF} - \overline{\Delta R}(T_{\rm tr}, \tfb) \right).
\end{equation}
Furthermore, the rate gap can be put in the general form (see \cite{Submitted})
\begin{equation} \label{g-gap}
\overline{\Delta R}(T_{\rm tr}, \tfb) = \log \left(1 + g(T_{\rm tr}, \tfb) \right)
\end{equation}
where the function $g(\cdot,\cdot)$ depends on the feedback strategy
and shall be specified later. Because the first multiplicative term is constant when $T_{\rm tr} + \tfb = T_t$ ,
the inner maximization corresponds to minimization of the function $g(\cdot,\cdot)$,
subject to the constraint $T_{\rm tr} + \tfb \leq T_t$.
Letting $g(T_t) \triangleq \min_{T_{\rm tr} + \tfb \leq T_t} g(T_{\rm tr}, \tfb)$ denote the solution of the inner maximization in (\ref{eq-mainopt2step}), we can solve the outer maximization by searching for the optimal value $0 < T_t \leq T$.

\subsection{TDD with channel reciprocity} \label{sect:tdd}

When channel reciprocity holds, open-loop CSIT estimation can be obtained from
the uplink pilot symbols. In this case, the amount of uplink training can be optimized as a special case of (\ref{eq-mainopt2step}) where no CSIT feedback is used.\footnote{Note that a similar optimization is
considered in \cite{jose:csi}, although in that work analysis of this optimization is not performed.}
In \cite[Remark 4.2]{Submitted}, the rate gap for a TDD system that uses $T_{\rm tr}$
{\em uplink} training symbols is given by:
\begin{equation} \label{tdd-gap}
\overline{\Delta R} = \log \left(1 + \frac{\nt - 1}{T_{\rm tr}} \right)
\end{equation}
which corresponds to $g^{\rm tdd}(T_{\rm tr})=\frac{\theta_{\rm tr}}{T_{\rm tr}}$ with $\theta_{\rm tr}=\nt - 1$.
Plugging this into (\ref{eq-mainopt}), we maximize the net spectral efficiency given by
\begin{eqnarray}\label{OptimizeTr}
f(T_{\rm tr})=\left(1-\frac{T_{\rm tr}}{T}\right) \left[R^{\rm ZF}- \log\left(1+ \frac{\theta_{\rm tr}}{T_{\rm tr}}\right)\right].
\end{eqnarray}
Because $f(\cdot)$ is concave in $T_{\rm tr}$,  the optimal $T_{\rm tr}^{\star}$
can be found by numerically solving for $\frac{\partial f}{\partial T_{\rm tr}}=0$ where
\begin{eqnarray}\label{Gradient}
\frac{\partial f}{\partial T_{\rm tr}} & = &
\frac{\theta_{\rm tr}\left(1-\frac{T_{\rm tr}}{T} \right)}{T_{\rm tr}^2 \left(1+ \frac{\theta_{\rm tr}}{T_{\rm tr}}\right)}
-\frac{1}{T} \left[R^{\rm ZF} - \log\left(1+ \frac{\theta_{\rm tr}}{T_{\rm tr}}\right)\right].
\end{eqnarray}
Although a closed-form solution for $T_{\rm tr}^\star$ cannot be found, we can study the scaling of the optimal $T_{\rm tr}^\star$
with the system parameters. It is not difficult to see that the derivative in (\ref{Gradient})
is upperbounded  by $\frac{1}{T} \widetilde{f}(T_{\rm tr})$, where
\begin{equation} \label{gradient-upperb}
\widetilde{f}(T_{\rm tr}) = \frac{\theta_{\rm tr}\left(T - T_{\rm tr}\right)}{T_{\rm tr}^2} - \left[ R^{\rm ZF} - \frac{\theta_{\rm tr}}{T_{\rm tr}} \right]
\end{equation}
The concavity of $f(\cdot)$ implies that the solution
$\widetilde{T}_t$ of the equation  $\widetilde{f}(T_{\rm tr}) = 0$ is an upper bound to the optimal value
$T^\star_t$. Solving for $\widetilde{f}(T_{\rm tr}) = 0$, we find
\begin{eqnarray}\label{TrScale}
T_{\rm tr}^{\star}  \leq \widetilde{T}_{\rm tr} = \sqrt{\frac{\theta_{\rm tr} T}{R^{\rm ZF}}}.
\end{eqnarray}
Furthermore, when the rate gap is small such that $\log\left(1+ \frac{\theta_{\rm tr}}{T_{\rm tr}}\right) \approx \frac{\theta_{\rm tr}}{T_{\rm tr}}$
(which becomes accurate for large $T$), the upperbound also becomes a very good
approximation.

Two interesting behaviors are obtained from  (\ref{TrScale}): 1) for a fixed SNR (i.e.,  constant $R^{\rm ZF}$)
$T^{\star}_{\rm tr}$ increases as $O(\sqrt{T})$ as $T \rightarrow \infty$;
2) for a fixed block length $T$,
$T^{\star}_{\rm tr}$ decreases as $O(1/\sqrt{R^{\rm ZF}})$ for large SNR, or equivalently, it decreases
as $O(1/\sqrt{\log(\rho)})$ since $R^{\rm ZF} = \log(\rho)+O(1)$ for large SNR.

Next, we examine the impact of $T_{\rm tr}^{\star}$ on the net achievable rate. By the definition of $T_{\rm tr}^\star$ we have:
\begin{align}\label{ApproxObj}
f \left( T_{\rm tr}^\star \right) &\geq f \left(\widetilde{T}_{\rm tr} \right)
 = \left(1 - \sqrt{\frac{\theta_{\rm tr}}{R^{\rm ZF} T}}\right) \left[R^{\rm ZF} - \log \LB 1+\sqrt{\frac{ \theta_{\rm tr} R^{\rm
ZF}}{T}}\RB \right]
\end{align}
The rate gap with respect to $R^{\rm ZF}$ can therefore be upper bounded as:
\begin{align}
R^{\rm ZF} - f \left(T_{\rm tr}^{\star} \right) &\leq  R^{\rm ZF} -  f \left(\widetilde{T}_t \right) \\
&= \sqrt{\frac{\theta_{\rm tr} R^{\rm ZF}}{T}}  + \log \LB 1+\sqrt{\frac{ \theta_{\rm tr} R^{\rm ZF}}{T}}\RB
- \sqrt{\frac{\theta_{\rm tr}}{R^{\rm ZF} T}} \log \LB 1+\sqrt{\frac{ \theta_{\rm tr} R^{\rm ZF}}{T}}\RB \label{temp2}\\ \label{EffectiveGap}
&\leq 2 \sqrt{\frac{\theta_{\rm tr} R^{\rm ZF}}{T}}
\end{align}
where the final inequality is reached by dropping the last term in (\ref{temp2}) and using
$\log(1+x) \leq x$. Thus, the gap to a perfect CSIT system decreases roughly as
$O(1/\sqrt{T})$ as $T$ increases.

For a future reference, it is worthwhile to notice that model 1 (a)
corresponds to model 1 (b) with perfect feedback such that the BS knows the UT channel estimates.
As a result, the net rate achievable with TDD, channel reciprocity and open-loop CSIT estimation
serves as an upper bound to the rate achievable with {\em any} form of CSIT feedback considered in the following.

\subsection{Analog Feedback}\label{sect:analog}

An option for the CSIT feedback scheme consists of sending the channel coefficients as QAM unquantized modulation symbols.
This is usually referred to as ``analog feedback'' in the literature, since the scheme is indeed akin to analog amplitude/phase
modulation. Because each UT is allowed $\frac{\tfb}{\nt}$ feedback channel uses,
this scheme transmits each channel coefficient over $\frac{\tfb}{\nt^2}$ feedback channel uses
(if $\tfb > \nt^2$, each coefficient is effectively repeated $\frac{\tfb}{\nt^2}$ times on the feedback channel).
At the BS receiver, MMSE estimation is used. The resulting rate gap is described as \cite[Section IV]{Submitted} and results in the $g(\cdot,\cdot)$ function
\begin{equation} \label{ganalog}
g^{\rm analog}(T_{\rm tr}, \tfb) = \frac{\nt - 1}{T_{\rm tr}} + \frac{\nt(\nt - 1)}{\tfb}.
\end{equation}
For the sake of generality, we consider a generalized form of (\ref{ganalog}) as
$g^{\rm analog}(T_{\rm tr}, \tfb) = \frac{\theta_{\rm tr}}{T_{\rm tr}} + \frac{\theta_{\rm fb}}{\tfb}$, for two non-negative weights $\theta_{\rm tr}$ and $\theta_{\rm fb}$. Comparing (\ref{ganalog}) with (\ref{tdd-gap}), we notice that the previous TDD open-loop case corresponds to letting $\theta_{\rm fb} = 0$, consistently with the fact that in this case no CSIT feedback is used.

It is immediate to check that the minimization of  $g^{\rm analog}(T_{\rm tr}, \tfb)$ subject to $T_{\rm tr} + \tfb = T_t$, and to $T_{\rm tr}, \tfb \geq 0$
is a convex problem. The corresponding Lagrangian \cite{boyd2004co} is given by
\[ \Lc(T_{\rm tr},\tfb,\mu) = g(T_{\rm tr}, \tfb)+\frac{1}{\mu^2} (T_{\rm tr} + \tfb) \]
where $\mu>0$ is the Lagrangian multiplier for the equality constraint. The KKT conditions \cite{boyd2004co} yield the solution
$T_{\rm tr}^{\star} = \sqrt{\theta_{\rm tr}} \mu$ and
$\tfb^{\star} = \sqrt{\theta_{\rm fb}} \mu$. Imposing the equality constraint and eliminating $\mu$, we obtain:
\begin{equation}\label{OptimalT}
T_{\rm tr}^{\star}  = \sqrt{\frac{\theta_{\rm tr}}{\Kc}} T_t, ~~~ T_{\rm fb}^{\star}  = \sqrt{\frac{\theta_{\rm fb}}{\Kc}} T_t
\end{equation}
where we let $\Kc = (\sqrt{\theta_{\rm tr}} + \sqrt{\theta_{\rm fb}})^2 $, and the resulting objective value
is given by $g^{\rm analog}(T_t)=  \frac{\Kc}{T_t}$.

The outer optimization (step 2) is now characterized in terms of a single variable $T_t$ and reduces to the maximization of (\ref{OptimizeTr}) where we replace $T_{\rm tr}$ and $\theta_{\rm tr}$ by $T_t$ and $\Kc$, respectively.
As a result, we find the optimal scaling for $T_t$ as
\begin{eqnarray}\label{TrScale}
T_{t}^{\star}  \leq \widetilde{T}_t = \sqrt{\frac{\Kc T}{R^{\rm ZF}}}.
\end{eqnarray}
Hence, the same analysis holds for the total length $T_t^{\star}$ of training and feedback. In addition, the following upper bound on $T_{\rm tr}^{\star}$ can be obtained by combining (\ref{TrScale}) with (\ref{OptimalT})
\begin{eqnarray}\label{T1}
T_{\rm tr}^{\star} \leq 
\sqrt{\frac{\theta_{\rm tr}}{\Kc}} \widetilde{T}_t = \sqrt{\frac{\theta_{\rm tr} T}{R^{\rm ZF}}} =
\sqrt{\frac{(\nt - 1) T}{R^{\rm ZF}}}.
\end{eqnarray}
According to this upperbound, the optimal downlink training is independent of $\theta_{\rm fb}$, and thus of the
efficiency of the feedback channel.

Similarly, we obtain the effective rate gap with respect to $R^{\rm ZF}$ as
\begin{align}
R^{\rm ZF} - f \left(T_{\rm tr}^{\star} \right)
&\leq 2 \sqrt{\frac{\Kc R^{\rm ZF}}{T}}
\end{align}
Comparing this and the corresponding expression (\ref{EffectiveGap}) for the open-loop TDD, we see that the analog feedback incurs a rate gap increase by a factor $1 + \sqrt{\nt}$.

\subsection{Error-Free Digital Feedback}\label{sec:dig-errorfree}

We now analyze a digital feedback technique where each UT quantizes its estimated channel vector into a $B$-bits message and then
maps these bits into $\frac{\tfb}{\nt}$ transmit symbols.
For the quantization step we consider an ensemble of random vector quantizers (RVQ) with directional
quantization as described in \cite{Jindal}. Assuming the feedback messages are received error-free,
in \cite[Section V]{Submitted} it is shown that the rate gap is given by
\begin{equation} \label{deltaR_digital}
\overline{\Delta R} = \log \left(1 + \frac{\nt - 1}{T_{\rm tr}} + \rho ~ 2^{-\frac{B}{\nt - 1}}  \right).
\end{equation}
For the time being, we assume unrealistically that error-free communication
is possible over the feedback channel at a rate equal to its capacity of
$\log_2 \left(1 + \rho \right)$ bits per channel use.  Letting  $B = \frac{\tfb}{\nt}\log_2 \left(1 + \rho \right)$, we obtain
\begin{equation} \label{g-digital}
g^{\rm digital}(T_{\rm tr}, \tfb) = \frac{\nt - 1}{T_{\rm tr}} + \rho \left(1 + \rho \right)^{- \frac{ \tfb }{\nt (\nt - 1)}}.
\end{equation}
Following the two-step approach, we minimize the above function subject to
$T_{\rm tr} + \tfb = T_t$. Since $g^{\rm digital}(\cdot,\cdot)$ is convex in $T_{\rm tr}, \tfb$,
we form the Lagrangian and readily obtain
\begin{eqnarray} \label{Tfb-errorfree}
T_{\rm tr} &=& \mu \sqrt{\nt - 1}, \;\; \tfb = \nt(\nt-1) \frac{ 2 \ln(\mu) + \ln \left(\frac{ \rho \ln(1+\rho) }{\nt (\nt-1)}\right)}{\ln (1 + \rho)}
\end{eqnarray}
where $\mu>0$ is chosen so that the equality constraint is fulfilled.
Note that $T_{\rm fb}$ grows as $O\left(\ln \mu\right)$,
much slower than the linear increase (in $\mu$) for $T_{\rm tr}$.

Contrary to the earlier analog feedback case, we cannot express $g^{\rm digital}(T_t)$ in a simple
closed form. However, using (\ref{Tfb-errorfree}) we can eliminate $\mu$ and
express $\tfb$ as a function of $T_{\rm tr}$:
\begin{eqnarray}
\tfb = \nt(\nt-1) \frac{ 2 \ln(T_{\rm tr}) + \ln \left(\frac{ \rho \ln(1+\rho) }{\nt (\nt-1)^2}\right)}{\ln (1 + \rho)},
\end{eqnarray}
and thus the net spectral efficiency can be written as:
\begin{eqnarray} \nonumber
& \left(1-\frac{T_{\rm tr} +  \nt(\nt-1) \frac{ 2 \ln(T_{\rm tr}) + \ln \left(\frac{ \rho \ln(1+\rho) }{\nt (\nt-1)^2}\right)}{\ln (1 + \rho)}}{T}\right) \times
    \left[R^{\rm ZF}- \log\left(1+ \frac{\nt - 1}{T_{\rm tr}} + \frac {\nt(\nt - 1)^2}{(T_{\rm tr})^2 \ln(1 + \rho)} \right) \right]. &
\end{eqnarray}
Because $\tfb$ increases logarithmically in $T_{\rm tr}$, and decreases with the SNR $\rho$,
its effect on the maximization is rather negligible. As a result, the maximization of $T_{\rm tr}$ is very similar to the case of
TDD with channel reciprocity. In other words, the error-free digital feedback performs almost as good as the TDD open-loop upper bound.

\subsection{Digital Feedback with Errors}\label{sec:dig-error}

We consider a practical digital feedback scheme with a very low complexity.
In particular, we assume that the $B$ feedback bits are transmitted on the uplink by using
uncoded QAM. Each UT makes use of $\frac{\tfb}{\nt}$ feedback channel uses for its CSIT feedback.
Assuming that quantization bits are arbitrarily mapped to the QAM constellation
symbols, the error of any symbol renders the feedback from a particular UT effectively useless
and thus leads to a zero rate.\footnote{This point can be made rigorous, but we limit ourselves to the present intuitive
argument for the sake of space limitation.}
Under this assumption, the achievable net spectral efficiency is given as a solution to
\begin{eqnarray}\label{errorDF-objective}
   \max_{T_{\rm tr}, \tfb: T_{\rm tr} + \tfb \leq T} \left(1-\frac{T_{\rm tr} + \tfb}{T}\right) (1-P_{e,\rm fb})
    \left[R_k^{\rm ZF} - \overline{\Delta R} \right]
\end{eqnarray}
where $\overline{\Delta R}$ is defined in (\ref{deltaR_digital}) and where $P_{e, {\rm fb}}$ is the feedback message error
probability. The size of the QAM constellation is given by $M = 2^{\frac{B \nt}{\tfb}}$ and yields a symbol error probability \cite{goldsmith2005wc}
\begin{equation}
P_s = 1 - \left(1 - 2 \left(1 - \frac{1}{\sqrt{M}}\right) Q \left( \sqrt{\frac{3\rho}{M-1}} \right) \right)^2,
\end{equation}
and a corresponding feedback message error probability
\begin{equation}
P_{e,\rm fb}  = 1 - (1 - P_s)^{\frac{\tfb}{\nt}}.
\end{equation}
Following the two-step optimization approach, we rewrite 
the outer optimization as
\begin{eqnarray}
 \max_{T_{\rm t}\leq T}\left(1-\frac{T_t}{T}\right)  \left[R^{\rm ZF} - \widetilde{\Delta R}(T_t) \right]
\end{eqnarray}
where the effective rate-loss $\widetilde{ \Delta R}(T_t)$, incorporating the loss due to erroneous feedback,
is given by
\begin{eqnarray} \label{minchia}
\widetilde{\Delta R}(T_t) & = & \min_{T_{\rm tr} + \tfb = T_t} \;\;
\left \{ \left(1 - P_{e,\rm fb} \right)
\log \left(1 + \frac{\nt - 1}{T_{\rm tr}} + \rho M^{-\frac{\tfb}{\nt(\nt - 1)}}  \right)
+ P_{e,\rm fb} R_k^{\rm ZF} \right \} .
\end{eqnarray}
If the QAM constellation size is suitably optimized,
the probability of feedback error can be made sufficiently small when the number of feedback bits $\frac{B \nt}{\tfb}$ per user
is large. For example, for $\nt=4$ at $10$ dB with $B = 25$ bits and 4-QAM, we have
$P_{e,\rm fb} = 0.0194$. As a result, the minimization in (\ref{minchia})
is very similar to the minimization of $g^{\rm digital}(T_{\rm tr}, \tfb)$ for error-free feedback in (\ref{g-digital}).

\begin{figure}
    \begin{center}
   \epsfxsize=4.0in
   \epsffile{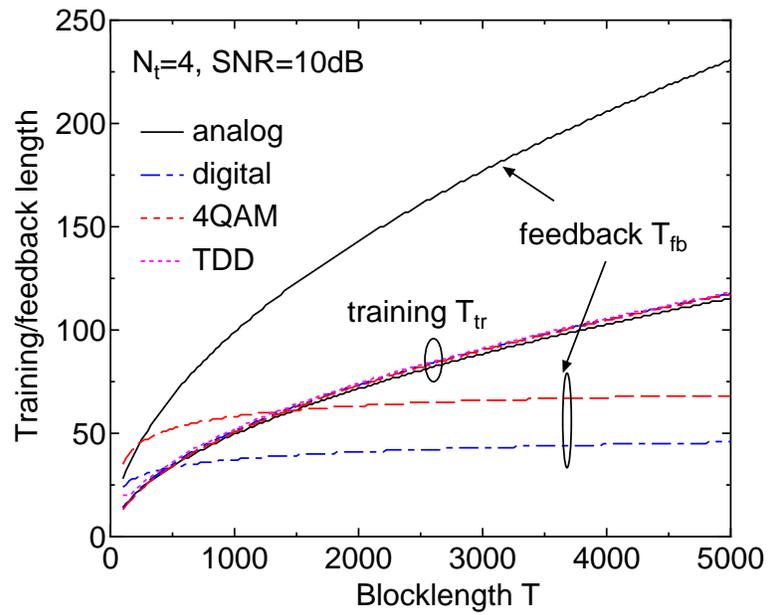}
    \end{center}
    \caption{Feedback/training length vs. block length for $N_t=4, \rho = 10$ dB.}
    \label{fig:TfbvsT}
\end{figure}

We conclude this section by providing some numerical examples to compare the performance of different feedback strategies. In Fig. \ref{fig:TfbvsT} the optimal values of $T_{\rm tr}$ and $\tfb$ are plotted versus
block length $T$ for analog feedback, error-free digital feedback, and QAM-based digital feedback along with the uplink training length $T_{\rm tr}$ for the TDD system.
Most striking is the fact that the optimal values of $T_{\rm tr}$
are essentially identical for the three feedback techniques as well as for TDD.
Furthermore, although not shown here, the optimal values of $T_{\rm tr}$ are
very well approximated by $\sqrt{\frac{(\nt - 1) T}{R^{\rm ZF}}}$ as in (\ref{T1}).
The number of feedback symbols, however, depends critically on the feedback method.
Because analog feedback is so inefficient, a large number of
feedback symbols are used so that the rate gap due to feedback is minimized.
On the other hand, digital feedback is very efficient and a relatively small number of
feedback symbols is required.

\begin{figure}
    \begin{center}
   \epsfxsize=4.0in
   \epsffile{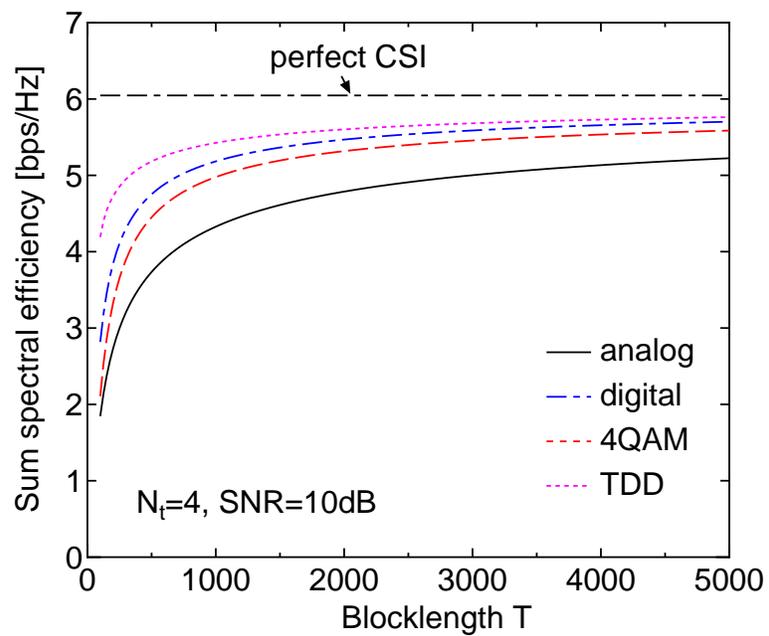}
    \end{center}
    \caption{Sum spectral efficiency vs. block length $T$.}
    \label{fig:RatevsT}
\end{figure}

In Fig. \ref{fig:RatevsT}, the sum spectral efficiency is plotted versus block length $T$. Although not shown here,  the rate approximations based upon (\ref{EffectiveGap}) are seen to become increasingly accurate as $T$ increases for analog
and TDD. Analog feedback is outperformed by digital feedback with or without errors, for any $T$. This is because
digital feedback offers a significantly smaller distortion as compared to analog
whenever $\tfb$ is larger than (approximately) $\nt^2$ (i.e., one symbol per channel coefficient)
\cite[Section VI]{Submitted}, and for reasonable block lengths it is optimal to use
$\tfb$ considerably larger than $\nt^2$ (see Fig. \ref{fig:TfbvsT}).

\section{Separate Uplink and Downlink Bandwidths} \label{sec:separate-bands}

In FDD systems, the uplink and downlink bandwidths are generally separated and the amount of channel uses per block length
dedicated to the CSIT feedback impacts the uplink spectral efficiency as an overhead,
rather than the downlink as in the previous section. In this section we focus on models 2 and 3 of Fig. \ref{fig:Model} assuming that the downlink and uplink bandwidths are a priori fixed.
The challenge here consists of determining the tradeoff region
of downlink spectral efficiency versus uplink CSIT feedback overhead.

For this purpose,  we consider the net downlink spectral efficiency, accounting for the training overhead,
as a function of $\tfb$.  For each value of $\tfb$, the optimal number of downlink training symbols is found, and the corresponding
net downlink spectral efficiency is given by:
\begin{align} \label{sukkia}
w\left(\tfb\right) &\triangleq \max_{T_{\rm tr} \leq T}  \;\;   \left(1-\frac{T_{\rm tr}}{T}\right) \left( R^{\rm ZF} - \log \left(1 + \frac{\nt-1}{T_{\rm tr}}+ \Delta(\tfb) \right) \right)
\end{align}
where $\Delta(\tfb)$ denotes the loss term due to CSIT feedback.
By solving for the maximization with respect to $T_{\rm tr}$, we obtain a tight lower bound on the
optimal downlink spectral efficiency achievable with ZF beamforming as a function of the parameter $\tfb$,
that quantifies the number of channel uses per block spent for the CSIT feedback over the uplink.

In the following, we first characterize such a tradeoff for the cases of the AWGN feedback channel based on model 2. Then, we address the case
of a temporally correlated channel with feedback delay and channel prediction by considering model 3.

\subsection{AWGN feedback link}

For the orthogonal access over the AWGN feedback channel, we have
$\Delta(\tfb)=\frac{N_t(N_t - 1)}{\tfb}$ for analog feedback, or  $\Delta(\tfb)=\rho \left(1 + \rho \right)^{- \frac{ \tfb }{\nt (\nt - 1)}}$ for error-free digital feedback
(see (\ref{ganalog}) and (\ref{g-digital})).
As seen previously, the effect of feedback errors can be made sufficiently small even by very simple schemes based on
uncoded QAM modulation. Hence, due to the space limitation, we provide only the analysis
for the case of error-free digital feedback operating at the uplink AWGN capacity, which captures the essential behavior of digital feedback while allowing for much
simpler analytical expressions. Nevertheless, in the numerical results we provide also
the results for a 4QAM-based digital feedback for the sake of comparison.

By simple manipulation, the objective function can be rewritten as:
\begin{align}
\left(1-\frac{T_{\rm tr}}{T}\right) \left( R^{\rm ZF} -  \log \left(1 + \Delta(\tfb) \right) -   \log \left(1 + \frac{N_t-1}{T_{\rm tr}(1 + \Delta(\tfb))} \right) \right).
\end{align}
Hence, the optimization has the same form as in Section \ref{sect:tdd},
with $R^{\rm ZF}$ replaced by $R^{\rm ZF} -  \log \left(1 + \Delta(\tfb) \right)$ and $\nt - 1$ replaced by $\frac{\nt - 1}{1 + \Delta(\tfb)}$.
It follows that we can immediately write the bound on the optimal training length as
\begin{equation}\label{ScalingAWGN}
T_{\rm tr}^{\star}(\tfb)  \leq \widetilde{T}_{\rm tr}(\tfb) = \sqrt{\frac{(\nt - 1) T}{\left(R^{\rm ZF} - \log \left(1 + \Delta(\tfb) \right)\right)
\left( 1 + \Delta(\tfb) \right)}}.
\end{equation}
Although $T_{\rm tr}^{\star}(\tfb)$ does depend on $\tfb$, this dependency is very weak whenever $\tfb$ is not too small. Thus,
very little is lost by simply choosing $T_{\rm tr} = \sqrt{\frac{(\nt - 1) T}{R^{\rm ZF}}}$.

Using the same arguments as in Section \ref{sect:tdd}, the downlink spectral efficiency can be lower bounded by
\begin{align}
w\left(\tfb\right) &\geq \left(1- \sqrt{\frac{\nt - 1}{T R^{\rm ZF}}}\right)
\left( R^{\rm ZF} - \log \left(1 + \sqrt{\frac{R^{\rm ZF}(\nt - 1)}{T}} + \Delta(\tfb) \right) \right) \\
&\geq R^{\rm ZF} - 2\sqrt{ \frac{R^{\rm ZF}(\nt - 1)}{T }} -  \left(\frac{1-\sqrt{\frac{\nt - 1}{T R^{\rm ZF}}}}
{1+\sqrt{\frac{R^{\rm ZF}(\nt - 1)} {T} }} \right) \Delta(\tfb).
\end{align}
Using the expressions for $\Delta(\tfb)$ we have:
\begin{align}\label{analograte-awgn}
w^{\textrm{analog}} \left(\tfb\right)
&\geq R^{\rm ZF} - 2\sqrt{ \frac{R^{\rm ZF}(\nt - 1)}{T }} -  \left(\frac{1-\sqrt{\frac{(\nt - 1)}{T R^{\rm ZF}}}}
{1+\sqrt{\frac{R^{\rm ZF}(\nt - 1)} {T} }} \right)  \frac{\nt(\nt-1)}{\tfb} \\ \label{digitalrate-awgn}
w^{\textrm{digital}} \left(\tfb\right)
&\geq R^{\rm ZF} - 2\sqrt{ \frac{R^{\rm ZF}(\nt - 1)}{T }} - \left(\frac{1-\sqrt{\frac{(\nt - 1)}{T R^{\rm ZF}}}}
{1+\sqrt{\frac{R^{\rm ZF}(\nt - 1)} {T} }} \right) \rho \left(1 + \rho \right)^{- \frac{ \tfb }{\nt (\nt - 1)}}.
\end{align}
Notice that the spectral efficiency penalties due to training and feedback are separable in these lower bounds.
Based upon these expressions, we expect that the downlink spectral efficiency $w^{\textrm{digital}}(\tfb)$ with digital feedback converges very quickly to the rate accounting for the optimized training overhead,
which is approximately $R^{\rm ZF} - 2\sqrt{ \frac{R^{\rm ZF}(\nt - 1)}{T }}$, whereas convergence is much slower with analog feedback.

The above definitions of $w^{\textrm{analog}}$ and $w^{\textrm{digital}}$ characterize the net downlink
spectral efficiency as a function of the number of uplink symbols per block length used for CSIT feedback.
In terms of system design, it is more meaningful to characterize the downlink \textit{rate} as a function of the uplink \textit{bandwidth}
used for channel feedback.
Under the block-fading model adopted in this paper, the channel is constant for $T_c$ seconds
over the bandwidth of $W_c$.
Since $\tfb$ uplink symbols are used for channel feedback for every block,
the uplink bandwidth used for channel feedback is given by $\frac{\tfb}{T_c}$ Hz, and the
downlink rate is given by $W_c w(\tfb)$ in bit/sec (bps).

We can take advantage of the above analysis to understand the fundamental tradeoff between downlink and uplink rate.
To this end, we employ a simplistic model of the uplink in which we assume
the uplink bandwidth of $W_{\rm  up}$
Hz and the uplink spectral efficiency of $C_{\rm  up}$ bps/Hz. Since feedback consumes $\frac{\tfb}{T_c}$
Hz of uplink bandwidth, the remaining bandwidth of $W_{\rm  up} -  \frac{\tfb}{T_c}$ Hz is available for uplink data
transmission. Thus the uplink data \textit{rate} is
\begin{equation}
R_{\rm {up}} (\tfb)=\left( W_{\rm  up} -  \frac{\tfb}{T_c} \right) C_{\rm  up}.
\end{equation}
while the downlink \textit{rate} is
\begin{equation}
R_{\rm {down}} (\tfb)= W_c w \left( \tfb \right).
\end{equation}
As $\tfb$ increases, the downlink rate $R_{\rm {down}}$ increases at the expense
of decreasing uplink rate $R_{\rm {up}}$. In order to determine the operating point on the $(R_{\rm {down}}, R_{\rm {up}})$ Pareto-optimal
boundary, a common method consists of maximizing the weighted sum of rates:
\begin{equation}
\max_{\tfb} ~~ \lambda R_{\rm {down}} (\tfb) + \overline{\lambda} R_{\rm {up}} (\tfb)
\end{equation}
where $0 < \lambda < 1$ and $\overline{\lambda} = 1 - \lambda$.
This optimization is equivalent to
\begin{equation}
\max_{\tfb} ~~ \lambda W_c w \left( \tfb \right) -  \overline{\lambda} \left(\frac{\tfb}{T_c} \right)  C_{\rm {up}}.
\end{equation}
After multiplying both sides by $T_c$ and taking the derivative with respect to $\tfb$, we see that the optimal solution satisfies:
\begin{equation}\label{Tfb-lambda}
\lambda T w' \left( \tfb \right) =  \overline{\lambda} C_{\rm {up}}  ~~ \rightarrow ~~
 w' \left( \tfb \right) =  \frac{1}{T} \frac{ \overline{\lambda}}{\lambda} C_{\rm {up}}.
\end{equation}
More precisely, we obtain the optimal $\tfb$ as a function of $\lambda$ as
\begin{eqnarray}\label{tfb-analog}
\tfb^{\rm analog}(\lambda) &=&  \sqrt{ \frac{rN_t(N_t-1)T\lambda}{C_{\rm up}\overline{\lambda}}}\\ \label{tfb-digital}
\tfb^{\rm digital}(\lambda) &=& \frac{\nt (\nt-1)}{\log(1+\rho)} \log\LB\frac{r \rho\log(1+\rho)T \lambda}{\nt (\nt-1)C_{\rm up} \overline{\lambda}}\RB
\end{eqnarray}
where we let $r=\frac{1-\sqrt{\frac{(\nt - 1)}{T R^{\rm ZF}}}} {1+\sqrt{\frac{R^{\rm ZF}(\nt - 1)} {T} }} $.
Clearly the feedback length is non-negative and upper bounded by $T$. Compared to analog feedback, the feedback length $\tfb^{\rm digital}(\lambda)$ with digital feedback is almost insensitive to $\lambda$ except the corner points ($\lambda=0,1$).
By plugging the above expressions into (\ref{analograte-awgn}), (\ref{digitalrate-awgn}),
the achievable rate can be parameterized by
$\lambda$ such that
\begin{align}
w^{\rm {analog}}(\lambda)
&\geq R^{\rm ZF} - 2\sqrt{ \frac{R^{\rm ZF}(\nt - 1)}{T }} - \sqrt{\frac{r\nt(\nt-1)C_{\rm up} \overline{\lambda}}{T\lambda}} \\
w^{\rm {digital}} (\lambda)
&\geq R^{\rm ZF} - 2\sqrt{ \frac{R^{\rm ZF}(\nt - 1)}{T }} -\frac{\nt(\nt-1)C_{\rm up} \overline{\lambda}}{T\lambda\log(1+\rho)} .
\end{align}
The third term, representing the rate loss due to the imperfect feedback, is rather marginal both for analog and digital feedback schemes
for a large $T$ in the range $0 < \lambda < 1$. From these expressions, it can be expected that the tradeoff curve with digital feedback is sharper and dominates the curve with analog feedback.

\begin{figure}
    \begin{center}
   \epsfxsize=4.0in
   \epsffile{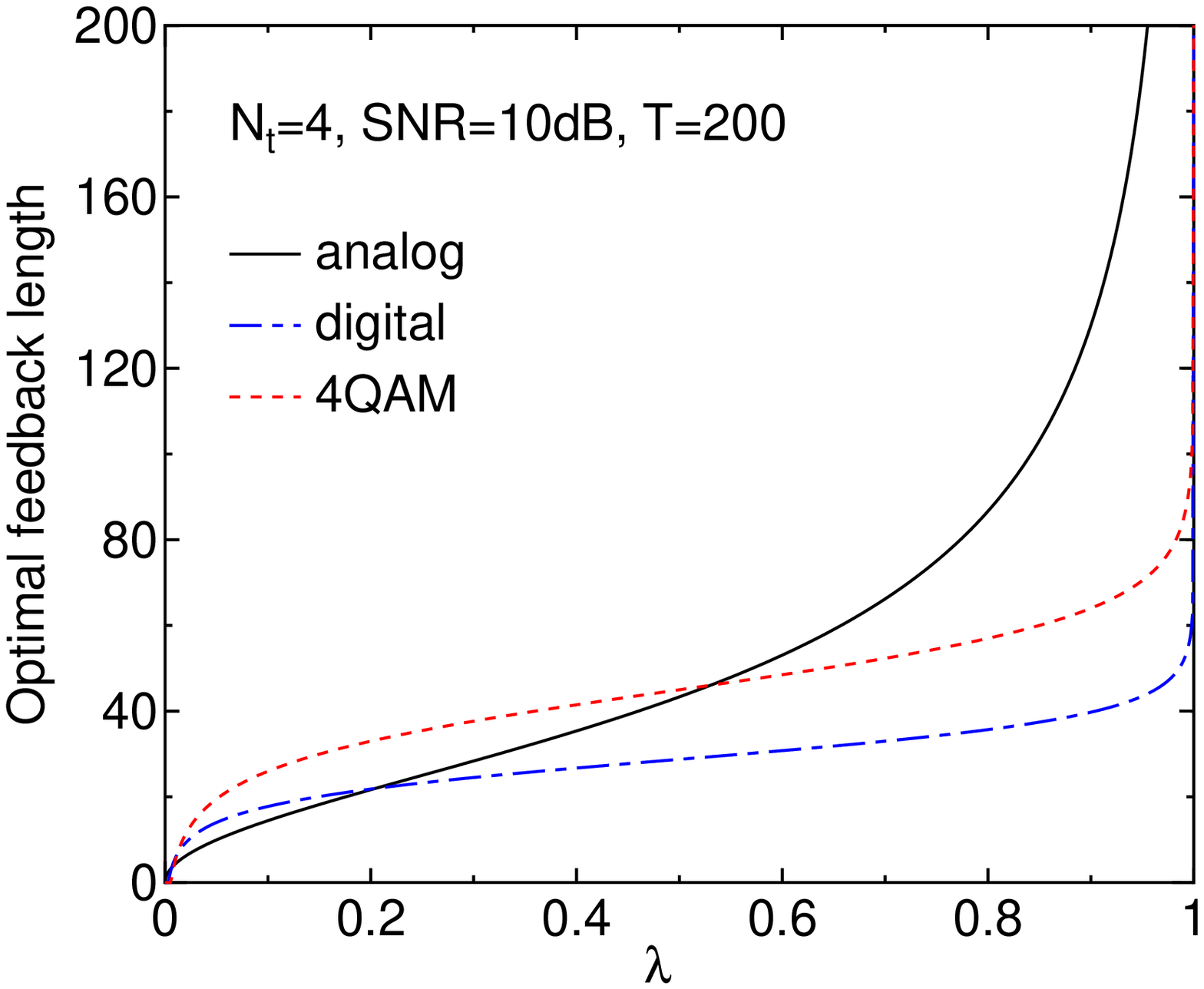}
    \end{center}
    \caption{Feedback length vs. $\lambda$ for $\nt=4$, $\rho = 10$ dB, $T=200$ symbols.}
    \label{fig:tfb-lambda}
\end{figure}

To make this discussion more concrete, consider a single resource block in LTE, with bandwidth $200$ kHz
and duration $1$ ms, corresponding to $T=200$ in our model. We assume $C_{\rm {up}}=1.512$ bps/Hz
(per user) and an uplink bandwidth also equal to $200$ kHz, for the sake of symmetry.
The uplink-downlink sum rate boundary (expressed in kbps) and the corresponding feedback
lengths are shown in Figs. \ref{fig:up_down} and \ref{fig:tfb-lambda}. A well-designed system will typically operate near the sharp ``knee" of the curves
of Fig. \ref{fig:up_down},
where the downlink rate is very close to its maximum value. Fortunately, because of the relatively low cost of channel feedback, the uplink rate is also reasonably close to its maximum. From Fig. \ref{fig:tfb-lambda} we remark also that analog feedback
requires a longer $\tfb$ for a larger weight $\lambda$ while the feedback length with digital feedback is almost constant. The training length was found to be 24 symbols for any scheme except for $\lambda\approx 0$.
Note that the choice $T=200$ is quite conservative.
As argued in Section \ref{sect:introduction}, typical physical channel parameters yield a significantly larger $T$ for
low mobility users.

The takeaway message of section is that, unless uplink data rate is very strongly preferred
over downlink data rate, it is efficient to operate the system at a point where the downlink spectral efficiency
is very close to the perfect-feedback case.

\subsection{Delayed feedback channel}

In this section we study the uplink/downlink tradeoff by taking into account the effect of the feedback
delay and the temporally correlated channel based on model 3. This model is motivated by the following scenario.
In practice, the downlink resource allocation blocks, i.e. the block bandwidth $W_f$ and block length $T_f$, might be defined a priori independently of $W_c$ and $T_c$, while these coherence parameters depend on the propagation environment as well as the users mobility and may even vary from user to user. For the case of a fixed block length $T=W_f T_f$ much shorter than $W_cT_c$, the channel coefficients in subsequent blocks are correlated.

In order to model such situation, we assume that the channel fading coefficients are
constant within each block of $T$ symbols and changes from block to block according to a stationary Gaussian random process with power
spectral density (Doppler spectrum) $S_h(\xi)$, strictly band-limited in $[-F,F]$, where $F < 1/2$ is the maximum normalized
Doppler frequency shift, given by $F = \frac{vf_c}{c} T_f$,
where $v$ is the mobile terminal speed (m/s), $f_c$ is the carrier frequency (Hz),
$c$ is the light speed (m/s). Furthermore, such a ``Doppler process'' satisfies
$\int_{-F}^F \log S_h(\xi) d\xi > -\infty$. This condition holds for most (if not all)
channel models usually adopted in the wireless mobile communication literature (see \cite{biglieri1998fci} and references therein),
where the Doppler spectrum has no spectral nulls within the support $[-F,F]$.
Because of symmetry and spatial independence, we can neglect the antenna index and consider scalar rather than vector processes.

Contrary to the block-by-block estimation previously considered,  each UT $k$ estimates $\hv_k(t)$ based on the observation $\{ s_k(t-\tau) : \tau = d, d+1, \dots, \infty\}$ available at UT $k$ up to block $t-d$
where $d$ denotes the feedback delay in blocks of length $W_f T_f$ and $s_k(t) = \sqrt{\frac{T_{\rm tr}P}{M}} h_k(t) + z_k(t)$ is the received signal
at UT $k$ at block $t$.
We focus on the case of $d=0$ (filtering) and $d=1$ (prediction) in the following.
The equivalent model for both cases is given by
\begin{eqnarray}\label{PredictionModel}
h_k(t) = \tilde{h}_k(t) + n_k(t)
\end{eqnarray}
where $\tilde{h}_k(t)=\EE[h_k(t)| \{ s_k(t-\tau)\}]$ denotes the estimated channel, independent of the estimation error
$n_k(t)\sim{\cal CN}(0,\sigma_{\rm tr}^2)$. The one-step prediction MMSE $(d=1)$ is given by \cite{lapidoth2005acs,Submitted}
\begin{eqnarray} \label{noisy-prediction}
\epsilon_1(\delta)
&=& \delta^{1 - 2F} \exp\left ( \int_{-F}^{F} \log(\delta + S_h(\xi)) d\xi \right ) - \delta
\end{eqnarray}
where we assume a unit-power process, $\int_{-F}^{F} S_h(\xi) d\xi = 1$, observed in background white noise with
per-component variance $\delta=\frac{\nt}{T_{\rm tr} \rho}$.
The filtering MMSE ($d=0$) is related to $\epsilon_1(\delta)$ through the well-known maximal ratio combining formula
\begin{eqnarray}\label{estimation-prediction}
\epsilon_0(\delta)
&=& \frac{\delta\epsilon_1(\delta) }{\delta + \epsilon_1(\delta)}.
\end{eqnarray}
Since $\tilde{h}_k(t)$ and $n_k(t)$ are independent, we have $\EE[|\tilde{h}_k(t)|^2] = 1 - \sigma_{\rm tr}^2$ for any $k$.

In \cite[Section VI. B]{Submitted}, it is shown that the rate gap is upper bounded by
\begin{eqnarray}\label{GapCorrelated}
\Delta R^{d} \leq 
\log\LB 1 + \frac{N_t-1}{T_{\rm tr}}\frac{\epsilon_d(\delta)}{\delta} + \Delta(\tfb) \RB.
\end{eqnarray}
For simplicity, we focus on the case of a uniform Doppler spectrum $S_h(\xi)=\frac{1}{2F}$ for $-F\leq \xi\leq F$.
This yields
\begin{eqnarray}\label{prediction-bound}
\frac{\epsilon_1(\delta)}{\delta} &=& \LB1 + \frac{1}{2F\delta}\RB^{2F} - 1 \leq \LB \frac{1}{2F\delta}\RB^{2F}
\end{eqnarray}
where the last inequality can be easily shown.
Using (\ref{estimation-prediction}) and (\ref{prediction-bound}) we obtain
\begin{eqnarray}
\frac{\epsilon_0(\delta)}{\delta} &\leq &\frac{1}{1+(2F\delta)^{2F}}.
\end{eqnarray}
Plugging these expressions into (\ref{GapCorrelated}), we obtain the rate gap upper bounds as
\begin{eqnarray}\label{filteringRateGapUB}
\overline{\Delta R}^{d=0} &=&  \log\LB 1 + \Delta(\tfb) +  \frac{\nt-1}{T_{\rm tr}}\frac{1}{1+\left(\frac{2F\nt}{\rho T_{\rm tr}}\right)^{2F}} \RB \leq  \log\LB 1 + \Delta(\tfb) +  \frac{\nt-1}{T_{\rm tr}} \RB \\
\overline{\Delta R}^{d=1} &= &  \log\LB 1 + \Delta(\tfb)+  \frac{\nt-1}{T_{\rm tr}} \LB \frac{\rho T_{\rm tr}}{2F\nt}\RB^{2F} \RB.
\end{eqnarray}
We observe that that for the case of filtering ($d=0$), the rate gap upper bound reduces
to that of the AWGN feedback link for sufficiently large $\rho$. In what follows, we consider the more interesting
case of one-step prediction.

We can again maximize the net downlink achievable spectral efficiency for the one-step prediction
case by solving
\begin{eqnarray}\label{Original}
w(T_{\rm fb}) &\eqdef &
    \max_{T_{\rm tr}\geq \nt} \left(1 - \frac{T_{\rm tr}}{T}\right) \left[R^{\rm ZF}(P)- \log\left(1+  \kappa T_{\rm tr}^{2F-1} + \Delta(T_{\rm fb})\right)\right]
\end{eqnarray}
where we defined the constant $\kappa = (\nt-1)\LB \frac{\rho}{2F\nt} \RB^{2F}$.
By letting the RHS of (\ref{Original}) denote $f(T_{\rm tr},T_{\rm fb})$,
we remark that the objective function $f(\cdot,\cdot)$ is concave in $T_{\rm tr}$. The the optimal $T_{\rm tr}$ in (\ref{Original})
satisfies
\begin{eqnarray}
\frac{\kappa (1-F)(T-T_{\rm tr})}{T_{\rm tr}^{2-F}(1+ \kappa  T_{\rm tr}^{1-F} + \Delta(T_{\rm fb}))} = R^{\rm ZF}-\log(1+\Delta(T_{\rm fb})) -
 \log\LB 1+ \frac{\kappa T_{\rm tr}^{-(1-F)}}{1+\Delta(T_{\rm fb})} \RB.
\end{eqnarray}
Following the same arguments as before, it follows that the solution $\widetilde{T}_{\rm tr}$ to the equation $\tilde{f}(T_{\rm tr}) = 0$
is an upper bound to the optimal $T^\star_{\rm tr}$, where
\begin{eqnarray}
\tilde{f}(T_{\rm tr})= \frac{\kappa (T-T_{\rm tr})}{T_{\rm tr}^{2-F}(1+\Delta(T_{\rm fb}))}- \left[R^{\rm ZF}-\log(1+\Delta(T_{\rm fb}))-\frac{\kappa
T_{\rm tr}^{-(1-F)}}{1+\Delta(T_{\rm fb})}  \right]
\end{eqnarray}
Explicitly, we find
\begin{eqnarray} \label{TrScale-bis}
T_{\rm tr}^{\star}(T_{\rm fb})  \leq \widetilde{T}_{\rm tr}(T_{\rm fb}) & = &
\LB\frac{(\nt-1) T}{(1+\Delta(T_{\rm fb}))\{R^{\rm ZF}-\log(1+\Delta(T_{\rm fb}))\}} \RB^{\frac{1}{2-F}} \LB \frac{\rho}{2FM} \RB^{\frac{2F}{2-F}}.
\end{eqnarray}
As $T$ increases, the training length $T_{\rm tr}$ scales as $O(T^{\frac{1}{2-F}})$ depending on the
Doppler frequency shift $0< F < \frac{1}{2}$. For a fixed $T$, the training length is increasing in $F$.
When the fading is quasi-static (i.e., very low mobility users with $v \approx 0$)
such that the channel becomes perfectly predictable, the training length coincides with the expression (\ref{ScalingAWGN}) for
the block-by-block estimation.
Since the term $\Delta(T_{\rm fb})$ is negligible for a sufficiently large $T_{\rm fb}$,
we can choose with little loss of optimality
\begin{equation}\label{predictionT1}
T_{\rm tr}=  \LB\frac{\kappa T}{R^{\rm ZF}} \RB^{\frac{1}{2-F}}=  \LB \frac{(\nt-1)T}{R^{\rm ZF}}\RB^{\frac{1}{2-F}}\LB \frac{\rho}{2F\nt} \RB^{\frac{2F}{2-F}}.
\end{equation}
Following in the footsteps of what has been done before,
we can obtain the lower bound of the downlink spectral efficiency as
\begin{eqnarray} \label{marimari}\nonumber
w(T_{\rm fb}) &\geq &  \left(1 - \frac{T_{\rm tr}}{T}\right) \left[R^{\rm ZF}- \log\left(1+  \kappa T_{\rm tr}^{2F-1} + \Delta(T_{\rm fb})\right)\right]  \\ \nonumber
& \geq & R^{\rm ZF}- \left[\frac{T_{\rm tr}}{T}R^{\rm ZF} +  (\nt-1)\LB \frac{\rho}{2F\nt} \RB^{2F}T_{\rm tr}^{2F-1}\right]- \frac{(1 - \frac{T_{\rm tr}}{T})\Delta(T_{\rm fb}) }{1+ (\nt-1)\LB \frac{\rho}{2F\nt} \RB^{2F} T_{\rm tr}^{2F-1}}\\
\end{eqnarray}
where we can replace $T_{\rm tr}$ by (\ref{predictionT1}).
Solving the weighted sum rate maximization, we obtain the optimal $\tfb$ in the same form of
(\ref{tfb-analog}) and (\ref{tfb-digital}),  for analog feedback and error-free digital feedback, respectively,
where the term $r$ is now replaced by $\frac{1 - \frac{T_{\rm tr}}{T}}{1+ (\nt-1)\LB \frac{\rho}{2F\nt} \RB^{2F} T_{\rm tr}^{2F-1}}$.

\begin{figure}
    \begin{center}
   \epsfxsize=4.0in
   \epsffile{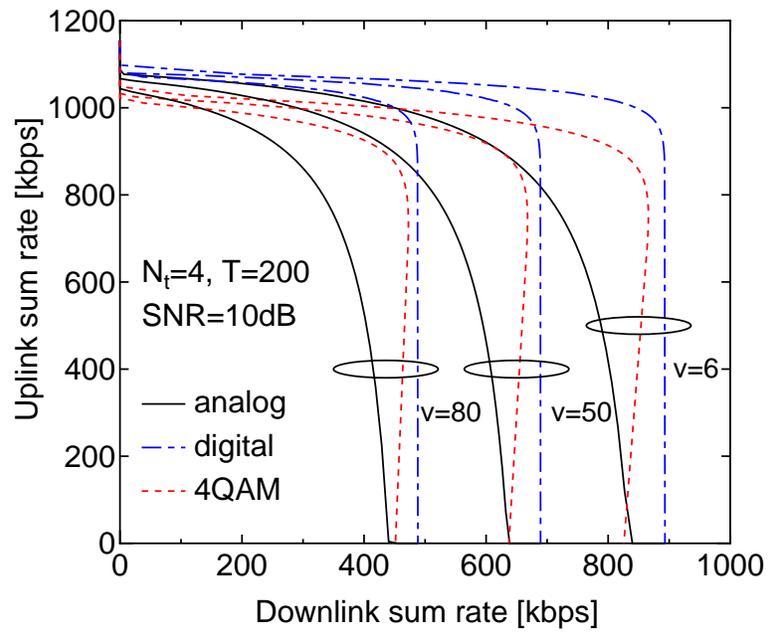}
    \end{center}
    \caption{Downlink vs. uplink tradeoff over the delayed feedback for $\nt=4$, $\rho = 10$ dB, $T=200$ symbols.}
    \label{fig:delaytradeoff}
\end{figure}

\begin{figure}
    \begin{center}
   \epsfxsize=4.0in
   \epsffile{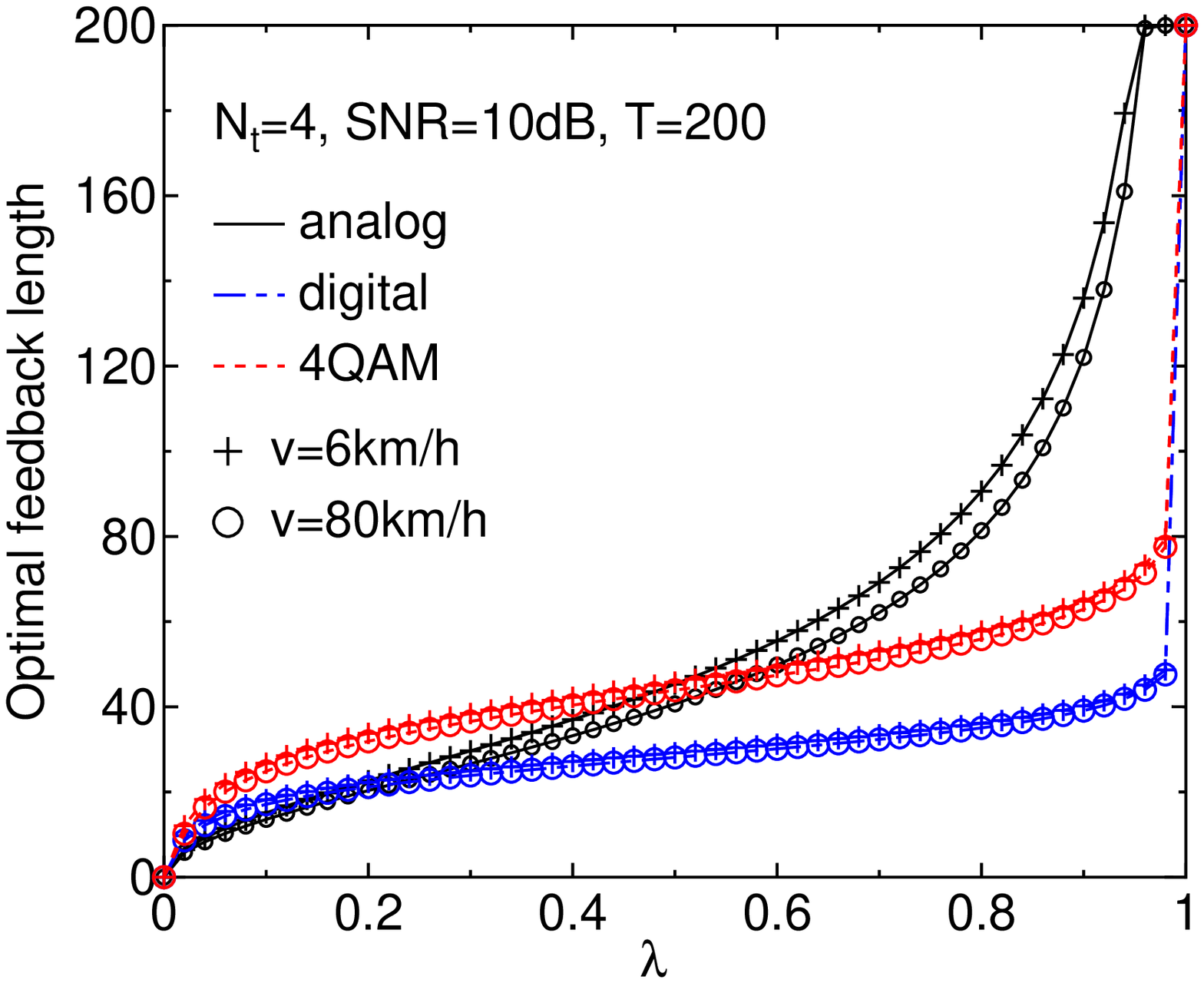}
    \end{center}
    \caption{Feedback length vs. $\lambda$ for $\nt=4$, $\rho = 10$ dB, $T=200$ symbols.}
    \label{fig:delaytfb-lambda}
\end{figure}

In order to quantify the impact of the delay on the uplink-downlink tradeoff, Fig. \ref{fig:delaytradeoff} shows the uplink-downlink sum rate
Pareto boundary for different mobile speeds $v=6, 50, 80$ km/h yielding the Doppler shift of $F=0.011, 0.093, 0.148$,
respectively, with the same parameters as Fig. \ref{fig:up_down}.
The corresponding feedback length as a function
of $\lambda$ is shown in Fig. \ref{fig:delaytfb-lambda}, where we only plotted for $v=6, 80$ km/h for the sake of clarity.
We recall that $\lambda=1$ corresponds to the corner point $(R_{\rm down},0)$
while $\lambda=0$ corresponds to the other corner point $(0,R_{\rm up})$.
As expected from (\ref{predictionT1}), the training length increases for a higher mobile speed and is found to be
$25, 36, 43$ symbols for $v=6, 50, 80$ km/h, respectively. On the contrary, the feedback length is rather indifferent to the mobile speed $v$,
although it tends to decrease for a larger $v$. On the uplink-downlink tradeoff curve, the higher mobile speed decreases significantly the downlink rate since the larger training length incurs a significant rate loss.

\begin{figure}
    \begin{center}
   \epsfxsize=4.0in
   \epsffile{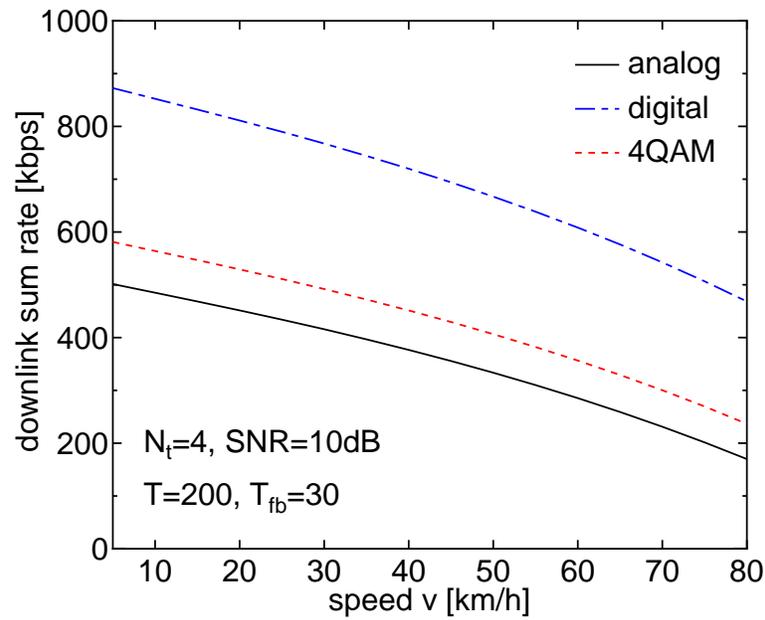}
    \end{center}
    \caption{Downlink rate vs. mobile speed for $\tfb=30$, $T=200$.}
    \label{fig:ratevsspeed-tfb20}
\end{figure}

Fig. \ref{fig:ratevsspeed-tfb20} shows the achievable downlink sum rate in kbps versus the mobile speed $v$ km/h when the uplink feedback length is set to $\tfb=30$ over a block length of $T=200$ symbols.
We compare analog feedback, error-free digital feedback as well as 4QAM-based digital feedback.
It is observed that by dedicating 15$\%$ of the uplink resource to the feedback, the uncoded 4QAM outperforms the analog feedback.

\section{Allowing for Many Users} \label{sec:many-users}

We conclude this paper by providing a discussion on the relevant case of $K > \nt$.
Until now we have assumed that the number of users is fixed equal to the number of BS antennas $N_t$.
In a real system there are often more than $N_t$ users (with data awaiting at the BS).
If more users feedback their channel information, the BS can utilize user selection
and generally obtain a non-negligible increase in downlink spectral efficiency.  Of course, allowing additional users to feed back
will incur a larger uplink bandwidth cost.  Indeed, a well designed system should optimize not only the total
number of feedback symbols used on the uplink, but also the number of users who feed back their channel state.
When the number of users enters into the picture, we see that the uplink-downlink tradeoff, which appeared rather trivial for a fixed
number of users, becomes indeed interesting and non-trivial.

Although the lower bound of \cite{Submitted} does not hold when user selection is performed, it can be numerically verified that
it is nonetheless a reasonable approximation of the rate with user selection and imperfect CSIT. For the sake of the space limitation, we focus on the separate uplink/downlink bands (model 2) although the other models can be adapted to the case of $K>\nt$ in a same manner. The corresponding downlink spectral efficiency is the solution to
\begin{align} \label{eq-separate_opt}
w\left(\tfb, K\right)
&\triangleq \max_{T_{\rm tr} : T_{\rm tr} \leq T}     \left(1-\frac{T_{\rm tr}}{T}\right) \left( R^{\rm ZF}_K - \log \left(1 + \frac{N_t - 1}{T_{\rm tr}} + \Delta(\tfb) \right) \right)
\end{align}
where now $R^{\rm ZF}_K$ denotes the perfect CSIT rate with ZF beamforming and user selection
\cite{yoo2006omb,DS05} and $K$ users. This is computed via Monte Carlo simulation due to
the lack of an analytical expression. Since the $\tfb$ feedback symbols are now split between $K$ users, we now have
$\Delta(\tfb)=\rho \left(1 + \rho \right)^{- \frac{ \tfb }{K (\nt - 1)}}$ for the case of error-free digital feedback.

\begin{figure}
    \begin{center}
   \epsfxsize=4.0in
   \epsffile{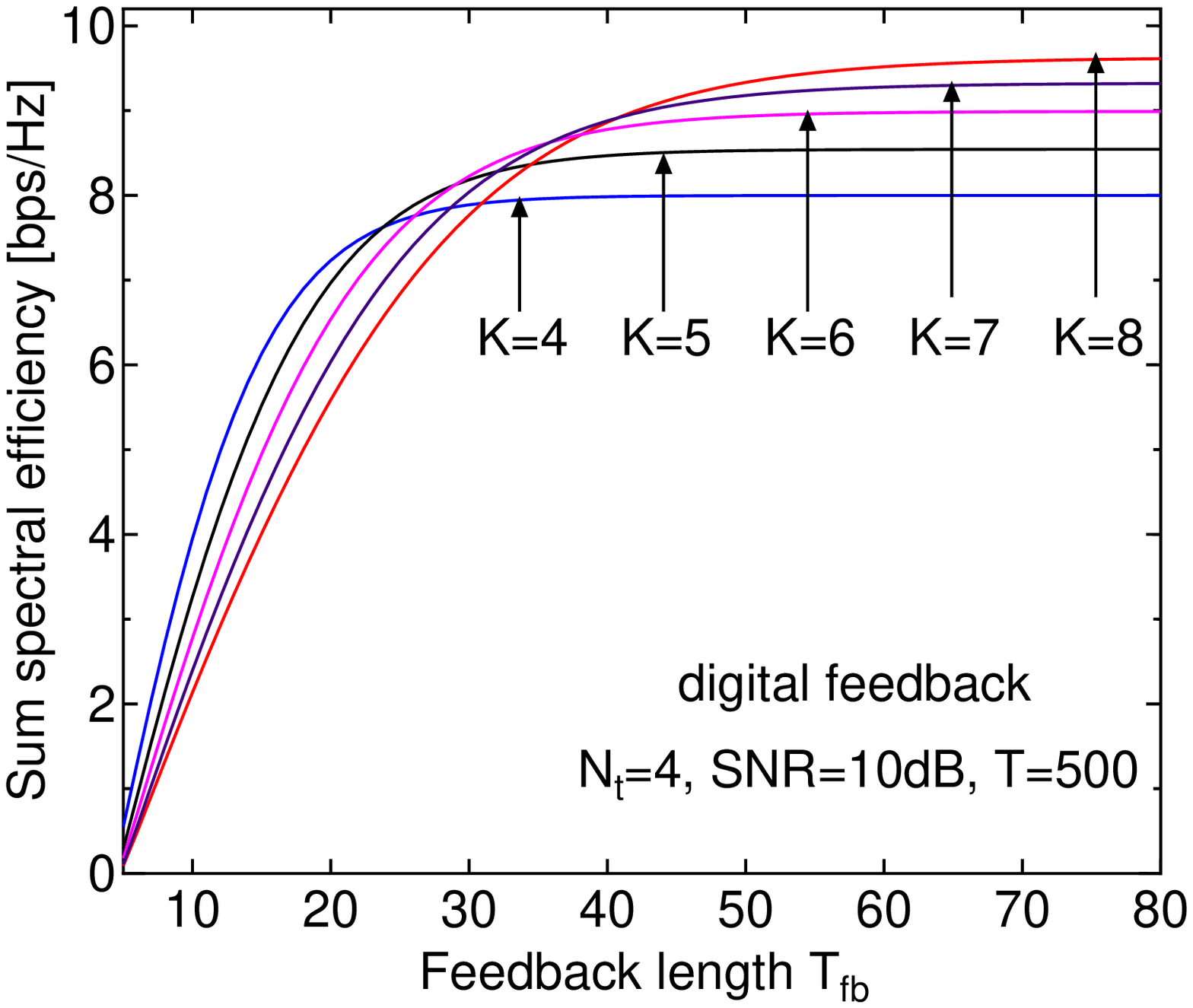}
    \end{center}
    \caption{Downlink sum spectral efficiency vs. feedback symbols ($\tfb$) for $T=500$, $\nt=4$, $\rho = 10$ dB, for $K$ from $4$ to $8$ users.}
    \label{fig:rate_vs_user}
\end{figure}

\begin{figure}
    \begin{center}
   \epsfxsize=4.0in
   \epsffile{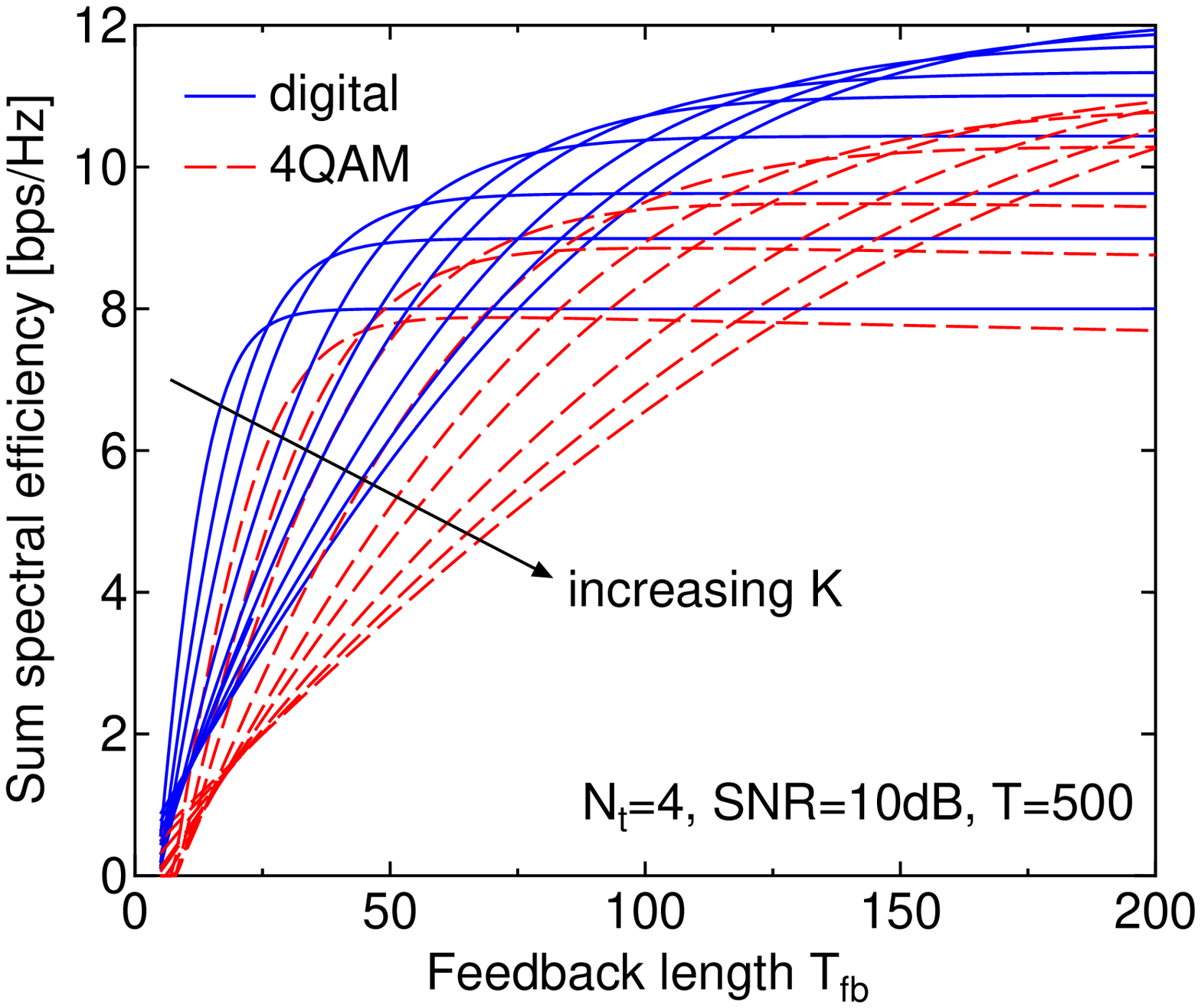}
    \end{center}
    \caption{Downlink sum spectral efficiency vs. feedback symbols ($\tfb$) for $T=500$, $\nt=4$, $\rho = 10$ dB, for $K$ from $4$ to $31$ users.}
    \label{fig:rate_vs_user2}
\end{figure}

In Fig. \ref{fig:rate_vs_user}, the downlink sum spectral efficiency $w\left(\tfb, K\right)$ is plotted versus $\tfb$ for $K=4,\ldots, 8$. The spectral efficiency is maximized
by letting $K=4, 5,6, 7, 8$ users feedback for $\tfb \leq 24$, $25\leq \tfb\leq 29$, $30\leq \tfb\leq 36$, $37\leq \tfb\leq 41$, $\tfb\geq 42$, respectively. Thus, the sum spectral efficiency is maximized by having approximately $\frac{\tfb}{6}$ users
feedback; this is very consistent with the findings of \cite{RavindranJindal07}.
If the number of users is fixed to $K=4$ there is virtually no benefit in increasing $\tfb$ beyond $35$ or $40$ because at that point the feedback channel is essentially perfect. However, a larger $\tfb$ enables more users to feed back and
yields a non-negligible gain in the achievable rate. For $\tfb \leq 200$, it turns out that no more than $31$ users are needed.  In Fig. \ref{fig:rate_vs_user2} the same plot
is given for $K=4,\ldots, 31$, for ideal digital and QAM feedback.   As $\tfb$ increases the marginal benefit of feedback (i.e., the slope)
decreases, but adding users does provide a reasonable benefit even up to the $31$-st user.

\begin{figure}
    \begin{center}
   \epsfxsize=4.0in
   \epsffile{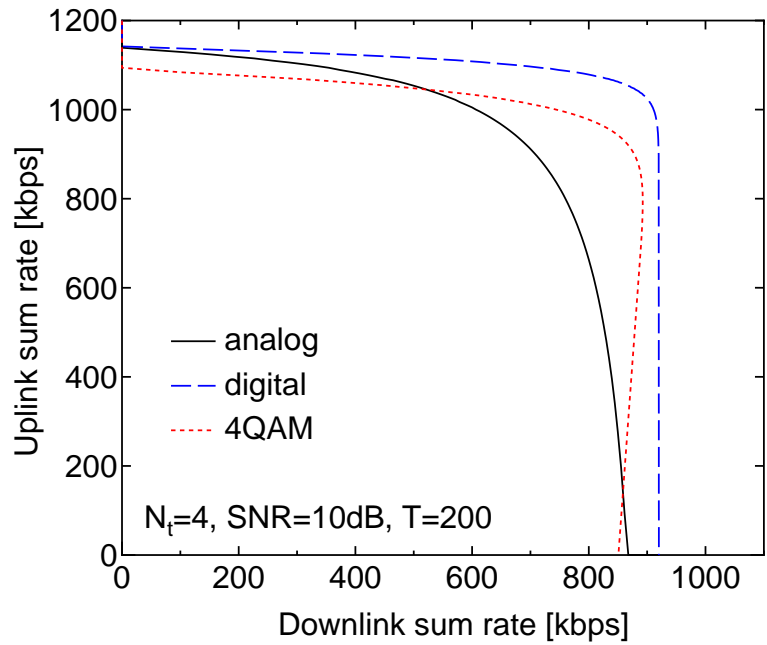}
    \end{center}
    \caption{Downlink vs. uplink tradeoff for $\nt=4$, $\rho = 10$ dB, $T=200$ symbols.}
    \label{fig:up_down}
\end{figure}

\begin{figure}
    \begin{center}
   \epsfxsize=4.0in
   \epsffile{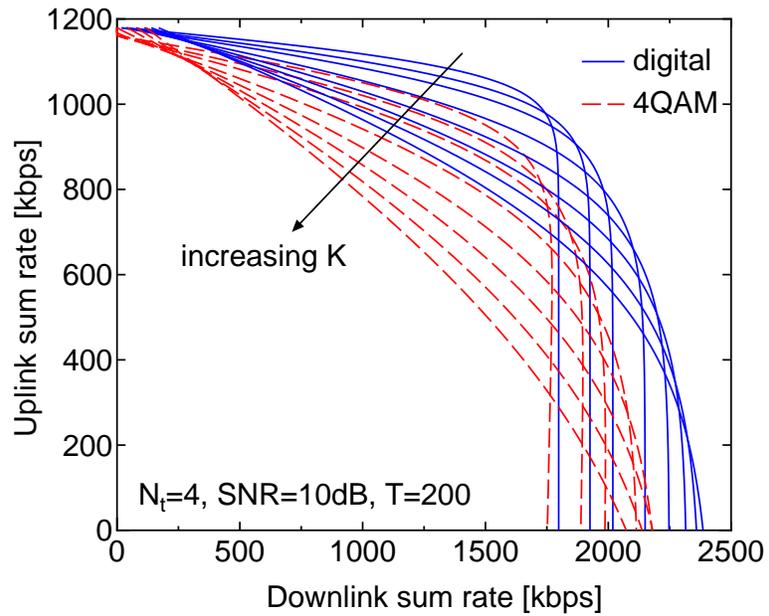}
    \end{center}
    \caption{Downlink vs. uplink tradeoff for $\nt=4$, $\rho = 10$ dB, $T=200$ symbols; allowing for up to $31$ users to feed back.}
    \label{fig:up_down2}
\end{figure}

We can also consider the tradeoff between uplink and downlink rate as done before.
Plotted in Fig. \ref{fig:up_down2} are the uplink and downlink sum rates, using precisely the same parameters as Fig. \ref{fig:up_down} (i.e., $T_c = 1$
msec and $W_c = 200$ kHz). We now see a non-trivial tradeoff for downlink rates larger
than $1750$ kbps (as before, it does not make sense to choose a smaller downlink rate than this unless uplink data rate is much more
strongly preferred than downlink data rate). If uplink and downlink data rates are equally weighted,
the optimal operating point corresponds to (approximately) $R_{\rm {up}} = 828$ Kbps and $R_{\rm {down}} = 1966$ Kbps, which is achieved with $K=11$ and $\tfb =63$ symbols. Note that the substantial benefit of allowing more users to feed back means that roughly 30 $\%$ of the uplink
bandwidth is used for channel feedback.

\bibliographystyle{IEEEtran}
\bibliography{asilomar06}

\newpage
\clearpage

\end{document}